\documentclass[preprint,showpacs,preprintnumbers,amsmath,amssymb]{revtex4}
\usepackage{graphicx}
\usepackage{bm}
\usepackage{color}

\begin{document}

\title{Impact of top-Higgs couplings on Di-Higgs production at future colliders}

\author{Ning Liu$^{1}$, Songlin Hu$^{1}$, Bingfang Yang$^{1,2}$ and Jinzhong Han$^{3}$}
\affiliation{$^1$College of Physics $\&$ Electronic Engineering, Henan Normal University, Xinxiang 453007, China\\
$^2$ School of Materials Science and Engineering, Henan Polytechnic University, Jiaozuo 454000, China\\
$^3$ School of Physics and Electromechnical Engineering, Zhoukou Normal University, Zhoukou, 466001, China \vspace*{1.5cm}  }

\begin{abstract}
Measuring the Higgs-self coupling is one of the most crucial goals of the future colliders, such as the LHC Run-II and the ILC-based photon collider. Since the new physics can affects the di-Higgs production not only from the Higgs self-coupling but also from the top-Higgs coupling, we investigate the di-Higgs production in the presence of the non-standard top-Higgs coupling at the LHC and ILC-based photon collider given the recent Higgs data. Due to the changed interference behaviors of the top quark loops with itself or $W$ boson loops, we find that the cross section of di-Higgs production at the LHC-14 TeV and ILC-500 GeV can be respectively enhanced up to nearly 3 and 2 times the SM predictions within 2$\sigma$ Higgs data allowed parameter region.

\end{abstract}

\maketitle
\section{Introduction}
In 2012, the ATLAS and CMS collaborations jointly announced that a bosonic resonance with a mass around 125 GeV was found at the LHC \cite{2012atlas,2012cms}. So far, most measurements of its properties are compatible with the predictions of the Higgs boson in the Standard Model (SM) \cite{2013atlas,2013cms}. However, due to the current limited statistics, the Higgs couplings with top quarks and with itself are still vacant and remain to be verified at the future colliders.

In the SM, the couplings of fermions to the Higgs boson are proportional to their masses. Due to the large mass, top quark has the strongest coupling to the Higgs boson and is speculated as a sensitive probe to the new flavor dynamics beyond the SM. As a direct way to test the top-Higgs coupling, the associated production of the top pair with Higgs boson has been widely studied at the LHC \cite{tth-atlas,tth-cms,tth-old,tth-new}. Besides, the search for a single top associated production with the Higgs boson was proposed to determine the sign of the top-Higgs coupling at the LHC \cite{thj-cms,thj-old,thj-new}. On the other hand, the top-Higgs coupling also plays a vital role in other processes involving the Higgs boson through the quantum effects, such as the di-Higgs production \cite{th-hh-lhc}. This makes the top-Higgs coupling inevitably entangled with the Higgs self-coupling and affects the measurement of the Higgs self-coupling at the LHC.

In the renormalizable Lagrangian of the SM, only the quartic Higgs coupling $\lambda(\phi^\dagger\phi)^2$ allowed by the electroweak gauge symmetry can generate the Higgs self-coupling. The measurement of the Higgs self-coupling is essential to reconstruct the Higgs potential and understand the electroweak symmetry breaking (EWSB) mechanism. In some extensions of the SM, the self-coupling can be significantly distorted by the loop corrections and become sensitive to the new physics \cite{3h-susy,3h-2hdm,3h-other}. Besides, a large deviation in the self-coupling may be a direct evidence for strong first-order electroweak phase transition in the early universe \cite{3h-cosmos}. At the LHC, the di-Higgs production is the only way to measure the Higgs self-coupling and is dominated by the gluon-gluon fusion mechanism, which has been widely studied in recent years \cite{hh-sm,hh-susy,hh-2hdm,hh-other}. Among various decay channels, although the $4b$ final state has the largest fraction, the rare process $hh \to b\bar{b}\gamma\gamma$ is expected to have the most promising sensitivity due to the low QCD backgrounds at the LHC \cite{hh-mc}.

In this work, we will investigate the effect of non-standard top-Higgs coupling in the di-Higgs production at the LHC and ILC-based photon collider under the current Higgs data constraints. Whenever examining the Higgs self-coupling at the LHC, one should keep in mind that, the main process $gg \to hh$ can also be triggered by the top-Higgs coupling itself through the box diagrams. These irrelevant processes lead to the strong cancellation with those involving the self-coupling in the SM, which makes the cross section of di-Higgs production nearly $10^3$ times smaller than the single Higgs production at the LHC. So, the top-Higgs coupling will affect the extraction of the Higgs self-coupling from the measurement of di-Higgs production at the LHC \cite{lilin}.

Given the limited precision of the LHC, an $e^+e^-$ collider is crucial to scrutinize the detailed properties of the Higgs boson that might uncover the new physics beyond the SM \cite{ilc-white}. In addition to the $e^+e^-$ collisions, high energy photon-photon collisions can be achieved at the ILC by converting the energetic electron beam to a photon beam through the backward Compton scattering \cite{ilc-rr}. Similar to the process $gg\to hh$ at the LHC, $\gamma\gamma \to hh$ also occurs at one-loop level. The measurements of the Higgs self-coupling at the photon collider were discussed in Ref. \cite{hh-rr-old}, where the complementarity of the photon and $e^+e^-$ collider was emphasized. Recently, an extensive study of the feasibility of the di-Higgs production with a parameter set of the ILC-based photon collider was reanalysed in Ref. \cite{hh-rr-new}, which concluded that that the channel $\gamma\gamma \to hh \to b\bar{b} b\bar{b}$ process can be observed with a statistical significance of about $5\sigma$ for the integrated luminosity corresponding to 5 years running of the photon collider. Therefore, the photon collider provide an ideal place to study the new physics effect in the di-Higgs production.

The structure of this paper is organized as follows. In Section \ref{section2}, we will briefly introduce the non-standard top-Higgs interaction and set up the calculations. In Section \ref{section3}, we present the numerical results and discuss the effects of non-standard top-Higgs coupling in the di-Higgs production at the LHC and ILC-based photon collider. Finally, we draw our conclusions in Section \ref{section4}.

\section{Top-Higgs interaction and Calculations\label{section2}}

In the SM the top-Higgs interaction can be written as:
\begin{eqnarray}
{\cal L}^{SM}_{t\bar{t}h}&=& -y_{t_{SM}} \overline{Q}_{3L}  t_R  \tilde\phi +\ {\rm h.c.},
\label{lyuk}
\end{eqnarray}
with
\begin{equation}
y_{t_{SM}}=\sqrt{2}m_t/v.\label{cyuk}
\end{equation}
where $Q_{3L}$ is the third generation SM quark doublet, $\phi$ is the Higgs doublet, $\tilde\phi_i =\epsilon_{ij}\phi_j$, and Higgs field vacuum expectation value (vev) $v\approx 246$~GeV. However, in some new physics models, the top-Higgs interaction can be different from the above SM prediction. These new physics effects on $t\bar{t}h$ coupling can be model-independently parameterized by a gauge invariant dimension-six operator \cite{eff}. For example, the term
\begin{eqnarray}
{\cal L}^{6}_{t\bar{t}h}&=& -\frac{C^{33}_{u\phi}}{\Lambda^2} (\phi^{\dag}\phi)(\overline{Q}_{3L} t_R \tilde\phi) +\ {\rm h.c.}.
\label{dim6}
\end{eqnarray}
Here we should note that the Eq.(\ref{dim6}) does correct the top quark mass $m_t$ by $\frac{v^3}{2\Lambda^2}[{\rm Re}\ C^{33}_{u\phi}+{\rm Im}\ C^{33}_{u\phi}\gamma_5]$, which has to be reabsorbed into the physical observable $m_t$ in Eq.(\ref{cyuk}). With this in mind, after the EWSB, we can have a general $t\bar{t}h$ interaction including the SM top-Higgs couplings and corrections from the dimension-six operator as following,
\begin{equation}
{\cal L}_{t\bar{t}h} = -\frac{y_{t}}{\sqrt{2}}\bar{t}(\cos\theta+i\sin\theta \gamma^{5})th,
\end{equation}
with
\begin{equation}
y_t \cos\theta=y_{t_{SM}}+\frac{v^2}{\Lambda^2}{\rm Re}\ C^{33}_{u\phi}, \quad y_t \sin\theta=\frac{v^2}{\Lambda^2}{\rm Im}\ C^{33}_{u\phi}.
\end{equation}
where $y_t$ takes the SM value $y_{t_{SM}}$ when ${\rm Re}\ C^{33}_{u\phi}={\rm Im}\ C^{33}_{u\phi}=0$. For convenience, we define two reduced couplings: $c_t= y_t \cos\theta/y_{t_{SM}}$ and $\tilde{c}_t= y_t \sin\theta/y_{t_{SM}}$ in the following calculations. Although the CP-violating interaction can contribute to the electric dipole moment (EDM), the bounds on the coupling $\tilde{c}_t$ rely on the assumption of Higgs couplings to other light fermions \cite{top_cp_review}. Given that these couplings are generally unobservable at the LHC, we do not impose EDM constraints in this study. For other low-energy physics constraints, such as $B_s-\bar{B}_s$ and $B \to X_s \gamma$, they are still too weak \cite{zupan}. The most relevant indirect constraint is from the current Higgs data since the non-standard top-Higgs interaction can change the production rate of $gg \to h$ and decay width of $h \to \gamma\gamma$ through the loop effect. The signal strengthes $\mu_i$ can be parameterized through the reduced couplings as following \cite{zupan},
\begin{eqnarray}
\mu_{hgg} \; & \simeq & c_t^2 + 2.6 {\tilde c}_t^2 + 0.11 c_t (c_t - 1) \, , \nonumber \\
\mu_{h\gamma\gamma} \; & \simeq & (1.28 - 0.28 c_t)^2 + (0.43 {\tilde c}_t )^2 \, .
\label{coulping}
\end{eqnarray}
We perform the $\chi^2$ fit of anomalous couplings $c_t$ and $\tilde{c}_t$ to the Higgs data by using the package \textsf{HiggsSignals-1.2.0} \cite{higgssignals}.

\subsection{$gg \to hh$}
\begin{figure}[th]
  \centering
  \includegraphics[width=16cm,height=6cm]{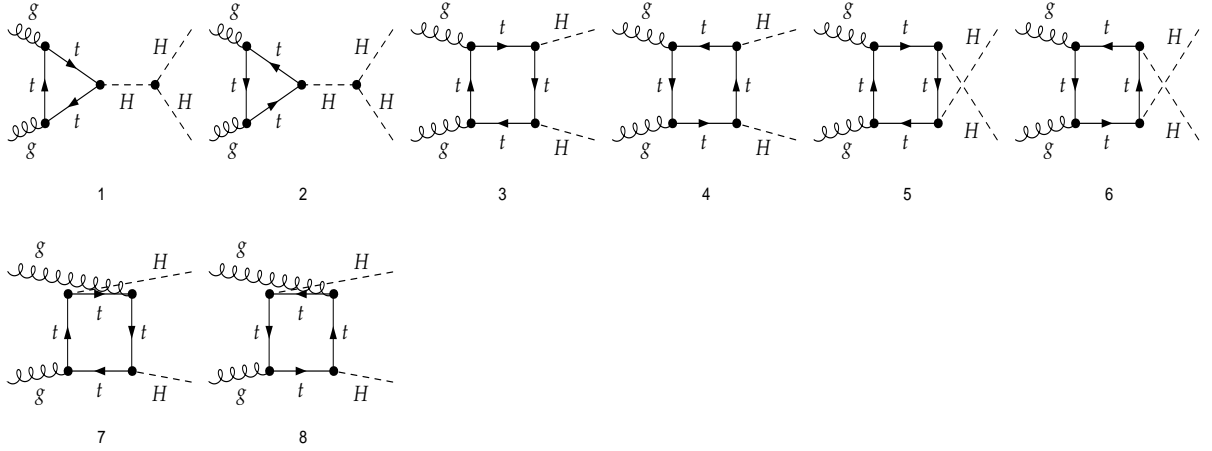}
  \caption{Feynman diagrams of the process $ gg\to hh$ at the LHC.}\label{fig1}
\end{figure}

In Fig.\ref{fig1}, we show the Feynman diagrams of the process $gg \to hh$ at the LHC. As above mentioned, the process $gg \to hh$ is generated by triangle
and box top quark loop diagrams, respectively. By applying the low energy theorem, we can obtain the effective coupling of any number of neutral scalar Higgs boson to two gluons \cite{hagiwara,spira},
\begin{equation}
{\cal L}_{hgg} = \frac{\alpha_s}{12 \pi}  G^a_{\mu\nu} G^{a\mu\nu} \log (1+\frac{h}{v})
= \frac{\alpha_s}{12 \pi}
\left(
\frac{h}{v} - \frac{h^2}{2v^2} + \cdots
\right)
G^a_{\mu\nu} G^{a\mu\nu}.\label{hgg}
\end{equation}
The first two interactions govern the cross section for di-Higgs production via the gluon fusion in the heavy top limit. From Eq.\ref{hgg}, we can see that there is a strong cancellation between the triangle and box top quark loops diagrams because of the opposite signs of the effective couplings. So if the new physics can flip the relative sign of them, the cross section of process $gg \to hh$ may be greatly enhanced.

\subsection{$\gamma\gamma \to hh$}
\begin{figure}[thpb]
\centering
\includegraphics[width=16cm,height=6cm]{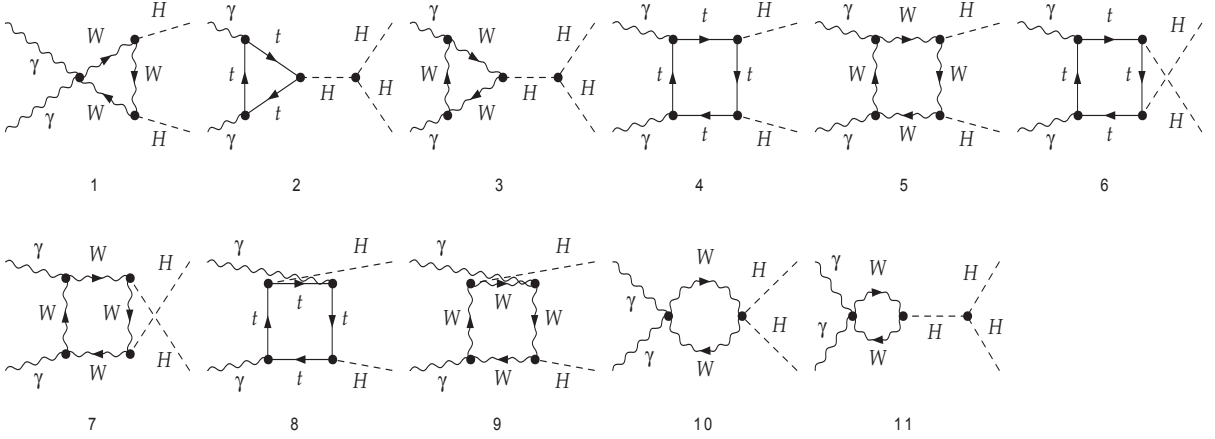}
\caption{Feynman diagrams of the process $\gamma \gamma \to hh$ at photon collider.}\label{fig4}
\end{figure}
In Fig.\ref{fig4}, we show the Feynman diagrams of the process $\gamma \gamma \to hh$, which is governed by $W$ boson and top quark loop diagrams, respectively. At the ILC, the $\gamma\gamma$ collisions are obtained by the inverse Compton scattering of the incident electron- and the laser-beam, the events number is calculated by convoluting the cross section of $\gamma\gamma$ collision with the photon beam luminosity distribution:
\begin{eqnarray}
N_{\gamma \gamma \to hh}&=&\int d\sqrt{s_{\gamma\gamma}}
  \frac{d\cal L_{\gamma\gamma}}{d\sqrt{s_{\gamma\gamma}}}
  \hat{\sigma}_{\gamma \gamma \to hh}(s_{\gamma\gamma})
  \equiv{\cal L}_{e^{+}e^{-}}\sigma_{\gamma \gamma \to hh}(s)
\end{eqnarray}
where $d{\cal L}_{\gamma\gamma}$/$d\sqrt{s}_{\gamma\gamma}$ is the photon-beam luminosity distribution and $\sigma_{\gamma \gamma \to hh}(s)$ ( $s$ is the squared center-of-mass energy of $e^{+}e^{-}$ collision) is defined as the effective cross section of $\gamma \gamma \to hh$, which can be written as \cite{photon collider}
\begin{eqnarray}
\sigma_{\gamma \gamma \to hh}(s)&=&
  \int_{\sqrt{a}}^{x_{max}}2zdz\hat{\sigma}_{\gamma \gamma \to hh}
  (s_{\gamma\gamma}=z^2s) \int_{z^{2/x_{max}}}^{x_{max}}\frac{dx}{x}
 F_{\gamma/e}(x)F_{\gamma/e}(\frac{z^{2}}{x})
\end{eqnarray}
where $F_{\gamma/e}$ denotes the energy spectrum of the back-scattered photon for the unpolarized initial electron and laser photon beams given by
\begin{eqnarray}
F_{\gamma/e}(x)&=&\frac{1}{D(\xi)}\left[1-x+\frac{1}{1-x}-\frac{4x}{\xi(1-x)}
  +\frac{4x^{2}}{\xi^{2}(1-x)^{2}}\right]
\end{eqnarray}
with
\begin{eqnarray}
D(\xi)&=&(1-\frac{4}{\xi}-\frac{8}{\xi^{2}})\ln(1+\xi)
  +\frac{1}{2}+\frac{8}{\xi}-\frac{1}{2(1+\xi)^{2}}.
\end{eqnarray}
Here $\xi=4E_{e}E_{0}/m_{e}^{2}$ ($E_{e}$ is the incident electron energy and $E_{0}$ is the initial laser photon energy) and $x=E/E_{0}$ with $E$ being the energy of the scattered photon moving along the initial electron direction. In the calculations, we fix the parameters as $\xi=4.8$, $D(\xi)=1.83$ and $x_{max}=0.83$ \cite{photon collider}.

Similar to the lagrangian ${\cal L}_{hgg}$, the effective coupling of any number of neutral scalar Higgs boson to two photons can be given as \cite{spira,shifman},
\begin{equation}
{\cal L}_{h\gamma\gamma} = \frac{\alpha_{em}}{2 \pi}(N_c\frac{Q^2_t}{3}-\frac{7}{4}) F^{\mu\nu} F_{\mu\nu} \log (1+\frac{h}{v})
= -\frac{47\alpha_{em}}{72 \pi}
\left(
\frac{h}{v} - \frac{h^2}{2v^2} + \cdots
\right)
F_{\mu\nu} F^{\mu\nu}.
\end{equation}
There is also the cancellation between the triangle and box loop diagrams. But different from $gg \to hh$, the contributions to the process $\gamma\gamma \to hh$ are dominated by the $W$ boson loops. So, the effect of the non-standard top-Higgs coupling on the di-Higgs production at photon collider may be smaller than that at the LHC.

For the loop calculations, we generate and simplify the amplitudes by using the packages \textsf{FeynArts-3.9} \cite{feynarts} and \textsf{FormCalc-8.2}\cite{formcalc}. All the loop functions are numerically calculated with the package \textsf{LoopTools-2.8} \cite{looptools}.

\section{numerical results and discussions \label{section3}}

In the numerical calculations, we take the input parameters of the SM as \cite{pdg}
\begin{eqnarray}
m_t=173.07{\rm ~GeV}, ~~m_W = 80.385~, ~~m_{Z}=91.19 {\rm
~GeV},\nonumber
\\m_h=125.9{\rm ~GeV},~~\sin^{2}\theta_W=0.2228, ~~\alpha(m_Z)^{-1}=127.918.
\end{eqnarray}
For the strong coupling constant $\alpha_s(\mu)$, we use its 2-loop evolution with QCD parameter $\Lambda^{n_{f}=5}=226{\rm ~MeV}$ and get $\alpha_s(m_Z)=0.118$. We use CTEQ6M parton distribution functions (PDF) for the calculation of $gg \to hh$ \cite{cteq}. The renormalization scale $\mu_R$ and factorization scale $\mu_F$ are chosen to be $\mu_R=\mu_F=m_h$. We numerically checked that all the UV divergences in the loop corrections canceled. Since the cross section of di-Higgs production is determined by the phase angle $\theta$ and the coupling $y_t$, we will firstly assume $y_t=y_{t_{SM}}$ and take $\theta=0,\pi/2,\pi$ for example to illustrate the effects of different phase angles on the di-Higgs production at the LHC and ILC-based photon collider in Fig.\ref{fig2} and Fig.\ref{fig5}, respectively. Then, we will vary both of $y_t$ and $\theta$ and respectively present the ratios of $\sigma^{gg \to hh}/\sigma^{gg \to hh}_{SM}$ and $\sigma^{\gamma\gamma \to hh}/\sigma^{\gamma\gamma \to hh}_{SM}$ under the constraint of the Higgs data in Fig.\ref{fig3} and Fig.\ref{fig6}. Here it should be noted that when $\theta\neq0$, the coupling $y_t$ with the SM value can be potentially dangerous because such a value may violate perturbative expansion of effective field theory and/or be inconsistent with the current LHC limit on the scale of new physics. So, in that case, $y_t < y_{t_{SM}}$ is usually needed to satisfy the theoretical and experimental bounds, which can be seen from Fig.\ref{fig3} and Fig.\ref{fig6}. For example, when $\theta=\pi/4$, $y_t/y_{t_{SM}}$ should be within the range  $0.4-0.6$.

\subsection{$gg \to hh$}

\begin{figure}[thp]
\centering
\includegraphics[width=9cm,height=7cm]{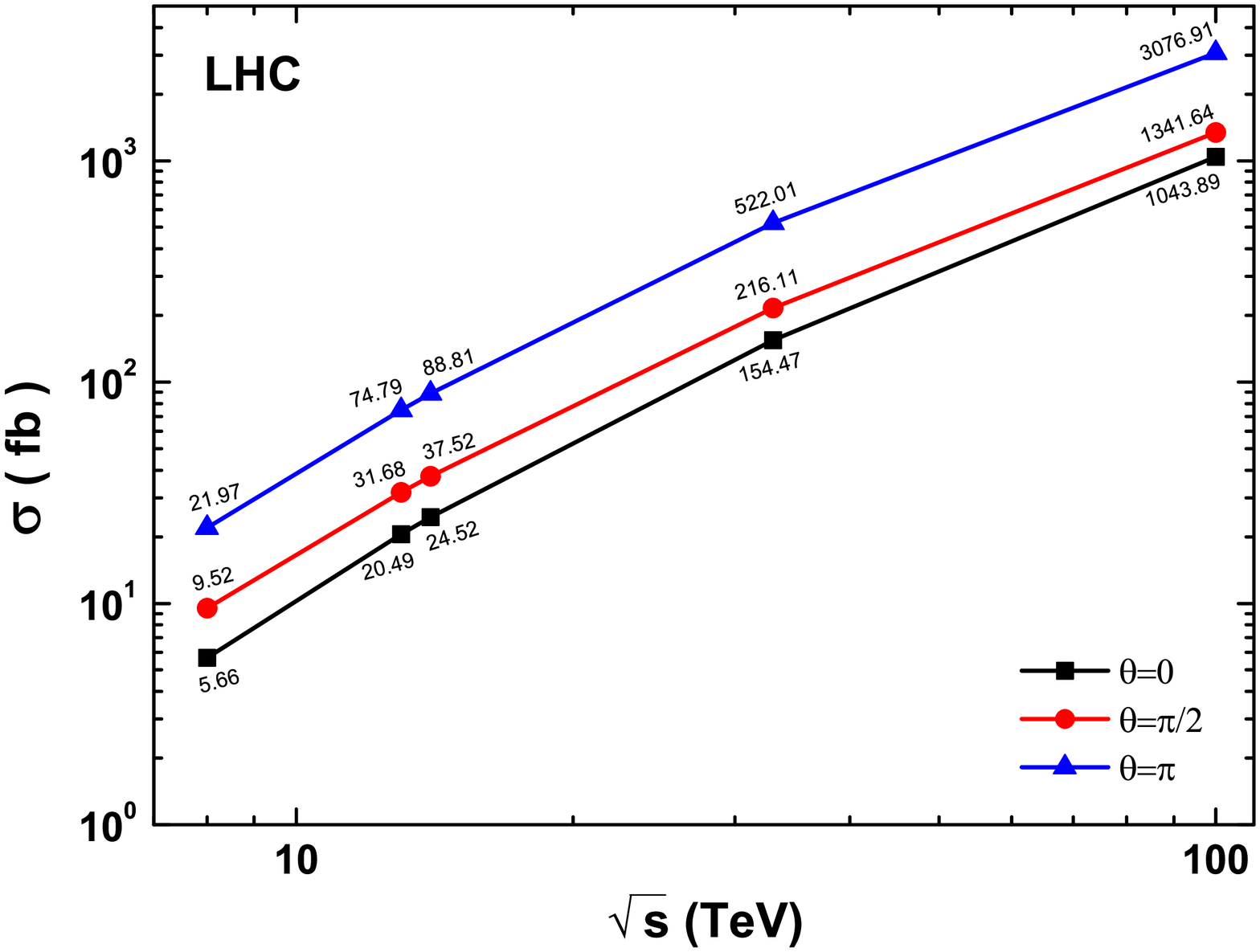}\vspace{-0.5cm}
\caption{Cross sections of the process $gg \to hh$ for non-standard top-Higgs couplings with $y_t=y_{t_{SM}}$ and $\theta=0,\pi/2,\pi$ at the LHC with $\sqrt{s}=8,13,14,33,100$ TeV.}\label{fig2}
\end{figure}
In Fig.\ref{fig2}, we present the impact of the non-standard top-Higgs couplings with $y_t=y_{t_{SM}}$ and $\theta=0,\pi/2,\pi$ on the cross section of the di-Higgs production at the LHC with $\sqrt{s}=8,13,14,33,100$ TeV. From Fig.\ref{fig2}, we can find that the top-Higgs coupling with an inverse sign $y_t=-y_{t_{SM}}$ can significantly enhance the di-Higgs cross sections from 24.52 fb to 88.61 fb, while the pseudo-scalar top-Higgs coupling can moderately increase the cross section up to 37.58 fb at $\sqrt{s}=14$ TeV. The reason is that these non-standard top-Higgs interactions can change the interference behavior of the triangle and box top loop diagrams in the process $gg \to hh$. To be specific, the amplitudes of these two kinds of Feynman diagrams for $\sqrt{s}\gg m_t, m_h$ in the SM can be approximately written as,
\begin{align}
{\cal M}^t_{box} & \sim  y_{t}^2\, \alpha_s \frac{m_t^2}{v^2}\, , \\
{\cal M}^t_{triangle} & \sim  -y_{t}\, \lambda_{hhh}\, \alpha_s \frac{m_t^2}{v^2} \, \frac{m_h^2}{\hat s}
\left[\log\left(\frac{m_t^2}{\hat s}\right) + i \pi\right]^2.
\end{align}
where we take the SM Higgs self-coupling $\lambda_{hhh}=3m^2_h/v$ in our study. For our cases, (i) when $\theta=\pi$, the top-Higgs coupling $y_{t_{SM}}$ becomes $-y_{t_{SM}}$ so that the sign of ${\cal M}_{triangle}$ is same as ${\cal M}_{box}$; (ii) when $\theta=\pi/2$, the SM coupling $y_{t_{SM}}$ changes to $iy_{t_{SM}}$. So there is no interference between the ${\cal M}_{triangle}$ and ${\cal M}_{box}$, which will provide a constructive contribution to $gg \to hh$.

Besides, we can see that the cross section of $gg \to hh$ becomes larger with the increase of $\sqrt{s}$ and can reach about 1 pb in the SM at $\sqrt{s}=100$ TeV, which is about 40 times larger than the one at $\sqrt{s}=14$ TeV. However, it should be noted that the amplitudes of triangle diagrams is suppressed by center of mass energy $\hat{s}$. The box diagrams will dominate the contribution to $gg \to hh$ at the VLHC. This means that the extraction of the Higgs self-coulping from the measurement of the total cross section of $gg \to hh$ will strongly depend on the assumption of the top-Higgs coupling at the VLHC. In this case, a study of the kinematic distributions of the Higgs bosons is needed to identify the sources of new physics in di-Higgs production.

\begin{figure}[thp]
\centering
\includegraphics[width=10cm,height=8cm]{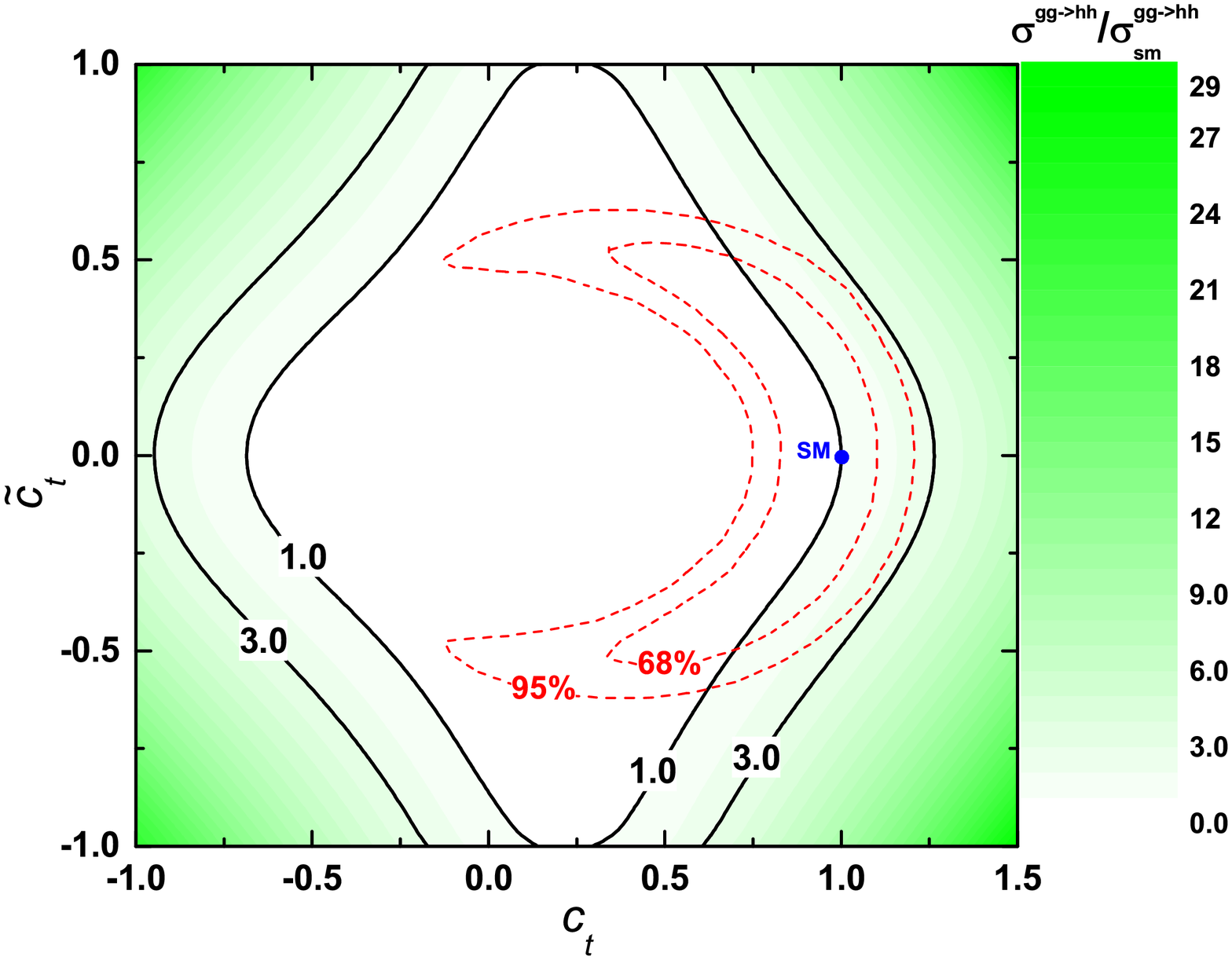}\vspace{0.5cm}
\vspace{-1cm} \caption{Ratios of $\sigma^{gg \to hh}/\sigma^{gg \to hh}_{SM}$ at 14 TeV in the plane of $\tilde{c}_t-c_t$, where the dashed contours correspond to the 68\% C.L. and 95\% C.L. limits given by the current Higgs data fitting.}\label{fig3}
\end{figure}
In Fig.\ref{fig3}, we plot the ratios of $\sigma^{gg \to hh}/\sigma^{gg \to hh}_{SM}$ at 14 TeV in the plane of $\tilde{c}_t-c_t$, where the dashed contours correspond to the 68\% C.L. and 95\% C.L. limits given by the current Higgs data fitting. From Fig.\ref{fig3}, we can see that the positive reduced scalar couplings $c_t > 0.5$ are strongly favoured and the reduced pseudoscalar couplings in the range $\tilde{c}_t > 0.6$ have been excluded at 95\% C.L by the Higgs data fitting. This severely constrains the enhancement of the di-Higgs production at the LHC. So the maximal value of the ratio $\sigma^{gg \to hh} /\sigma^{gg \to hh}_{SM}$ can only reach about 3 in the 95\% C.L. allowed region at 14 TeV LHC. On the other hand, the precise measurement of $gg \to hh$ will further bound these non-standard top-Higgs couplings.

\subsection{ ILC-based Photon Collider }

\begin{figure}[htpb]
\centering
\includegraphics[width=9cm,height=7cm]{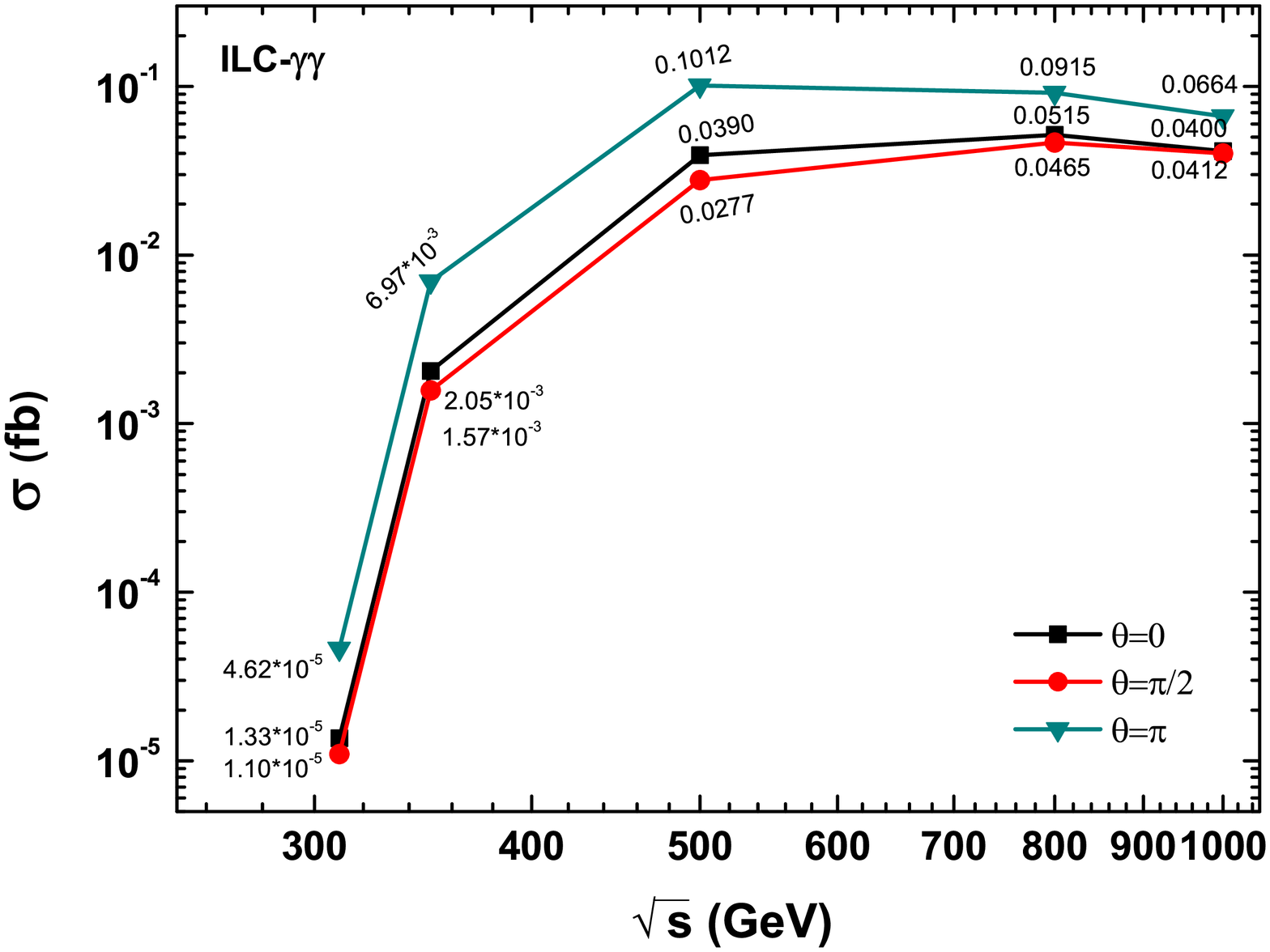}\vspace{-0.5cm}
\caption{Cross sections of $\gamma\gamma \to hh$ in the presence of the non-standard top-Higgs couplings with $y_t=y_{t_{SM}}$ and $\theta=0,\pi/2,\pi$ at the ILC-based photon collider with $\sqrt{s}=310,350,500,800,1000$ GeV.}\label{fig5}
\end{figure}
In Fig.\ref{fig5}, we show the cross sections of $\gamma\gamma \to hh$ in the presence of the non-standard top-Higgs couplings with $y_t=y_{t_{SM}}$ and $\theta=0,\pi/2,\pi$ at the ILC-based photon collider with $\sqrt{s}=310,350,500,800,1000$ GeV. From Fig.\ref{fig5}, we can see that the cross sections for pseudo-scalar coupling $\theta=\pi/2$ is smaller than that for the SM coupling $\theta=0$. This is different from the case of $gg \to hh$ at the LHC, where only top quark propagates in the loops. But for the process $\gamma\gamma \to hh$, $W$ boson loops will be involved and have an interference with the top quark loops. To be specific, the amplitudes of the top quark and $W$ boson box diagrams with $\sqrt{s}\gg m_t, m_W, m_h$ in the SM can be approximately written as,
\begin{align}
{\cal M}^t_{box} & \sim  y_{t}^2\, Q^2_t \alpha \frac{m_t^2}{v^2}\, , \\
{\cal M}^W_{box} & \sim  y_{W_{SM}}^2\, Q^2_W \alpha \frac{m_W^2}{v^2}.
\end{align}
where $Q_{t,W}$ is the electric charge of top quark and $W$ boson, respectively. $y_{W_{SM}}=gM_Z/\cos\theta_W$ denotes the SM Higgs gauge coupling and is fixed in our calculations. For $\theta=\pi/2$, the coefficient of the amplitude of the top quark box diagrams will be changed from $y^2_{t_{SM}}$ to $-y^2_{t_{SM}}$. So, as comparison with the SM prediction, the relative sign between the $W$ boson box and the top quark box will be inverted, which leads to a cancellation between them; For $\theta=\pi$, the flipped sign of $y_{t_{SM}}$ can increase the cross section of $\gamma\gamma \to hh$ in two sides: one is from the enhancement of those triangle diagrams involving $h\gamma\gamma$; the other one is from the constructive interference between the top quark triangle and box diagrams. We also note that for different non-standard top-Higgs couplings, the cross section of $\gamma\gamma$ always reach the maximal value at $\sqrt{s}=500$ GeV. This is caused by the threshold effect of the top quark pair in the loop. When $\sqrt{s}$ becomes larger, the cross section will decrease.

\begin{figure}[htpb]
\centering
\includegraphics[width=10cm,height=8cm]{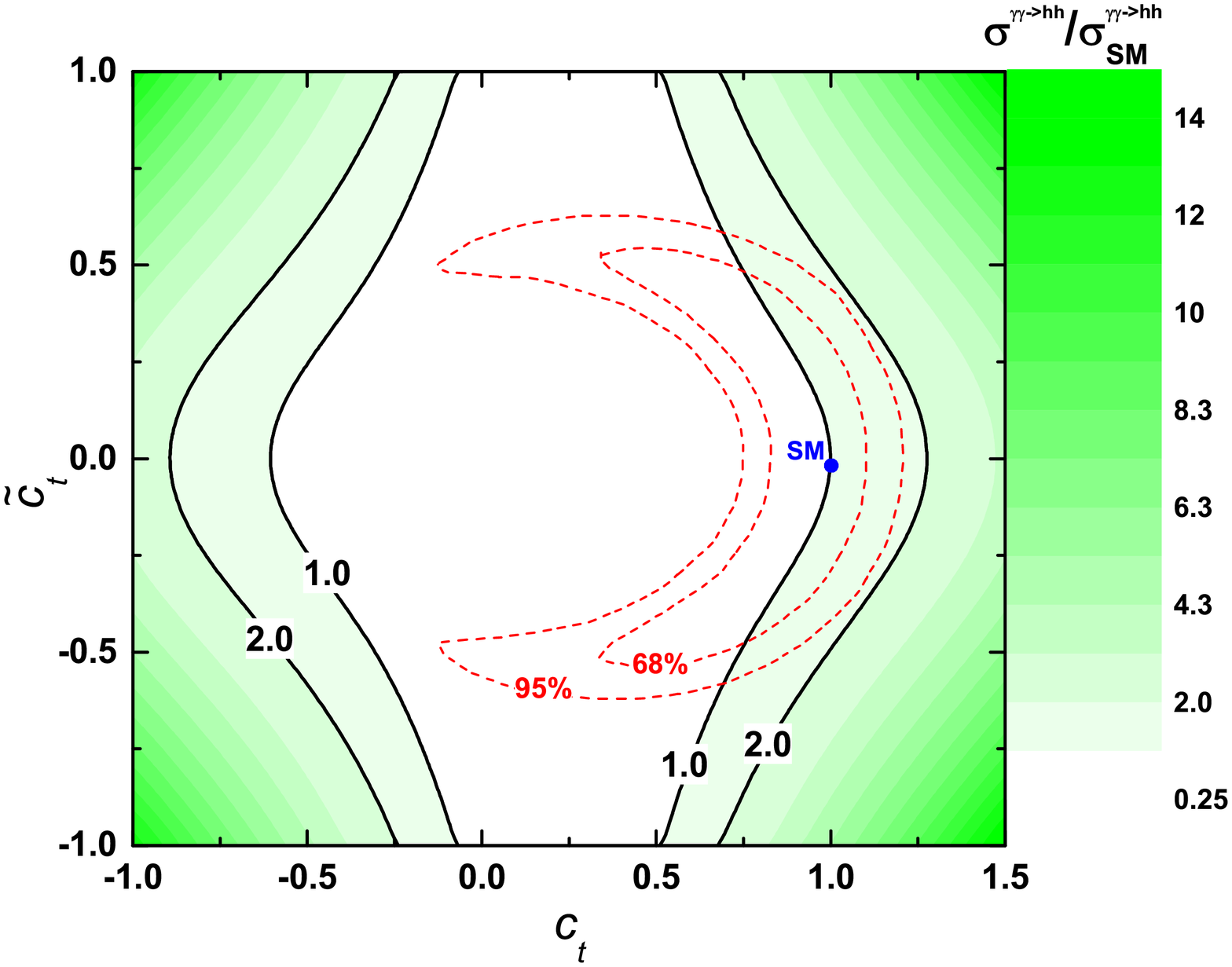}\vspace{0.5cm}
\vspace{-1cm} \caption{Similar to Fig.\ref{fig3}, but ratios of $\sigma^{\gamma\gamma \to hh}/\sigma^{\gamma\gamma \to hh}_{SM}$ at ILC-based photon collider with $\sqrt{s}=500$ GeV in the plane of $\tilde{c}_t-c_t$.}\label{fig6}
\end{figure}
Similar to Fig.\ref{fig3}, we plot the ratios of $\sigma^{\gamma\gamma \to hh}/\sigma^{\gamma\gamma \to hh}_{SM}$ at ILC-based photon collider with $\sqrt{s}=500$ GeV in the plane of $\tilde{c}_t-c_t$. From Fig.\ref{fig6}, we can see that although the cross section of $\sigma^{\gamma\gamma \to hh}$ can be about $13.9$ times the SM prediction, the maximal value of the ratio $\sigma^{\gamma\gamma \to hh} /\sigma^{\gamma\gamma \to hh}_{SM}$ can only reach about 2 in the region allowed by the current Higgs data at 95\% C.L.. So, given the latest analysis of the feasibility of $\gamma\gamma \to hh \to b\bar{b}b\bar{b}$, such an enhancement effect can be observed at the future photon collider.

\section{conclusions \label{section4}}
After LHC Run-I, measurement of Higgs self-coupling is one of the crucial tasks at future colliders, such as the LHC Run-II and the ILC-based photon collider. In this paper, given the recent Higgs data, we investigate the di-Higgs production in the presence of the non-standard top-Higgs coupling at the LHC and ILC-based photon collider. Due to the changed interference behaviors of the top quark loops with itself or $W$ boson loops, we find that the cross section of di-Higgs production at the LHC-14 TeV and ILC-500 GeV can be respectively enhanced up to nearly 3 and 2 times the SM predictions within 2$\sigma$ Higgs data allowed parameter region.

\section*{Acknowledgement}
This work is supported by the National Natural Science Foundation of China (NNSFC) under grants Nos. 11222548, 11275057, 11305049 and 11347140, by Specialised Research Fund for the Doctoral Program of Higher Education under Grant No. 20134104120002 and by the Startup Foundation for Doctors of Henan Normal University under contract No.11112.

\end{document}